\documentclass[aps,amsmath,graphicx,twocolumn]{revtex4}
\usepackage{graphics}
\usepackage{graphicx}
\usepackage{amsmath}
\usepackage{amssymb}
\usepackage{epsfig}

\usepackage[caption=false]{subfig}

\begin{document}

\title{Extension of the basis set of linearized augmented plane wave method (LAPW) by using supplemented tight binding basis functions}

\author{A.V. Nikolaev}
\email{alex_benik@yahoo.com}
\affiliation{Skobeltsyn Institute of Nuclear Physics Lomonosov Moscow State University, Leninskie gory, Moscow 119991, Russia}
\affiliation{Moscow Institute of Physics and Technology, 141700 Dolgoprudny, Russia}

\author{D. Lamoen}
\affiliation{EMAT, Department of Physics, Universiteit Antwerpen,
Groenenborgerlaan 171, 2020 Antwerpen, Belgium}

\author{B. Partoens}
\affiliation{CMT group, Department of Physics, Universiteit
Antwerpen, Groenenborgerlaan 171, 2020 Antwerpen, Belgium}

\begin{abstract}
In order to increase the accuracy of the linearized augmented plane wave method (LAPW) we present a new approach
where the plane wave basis function is augmented by two different
atomic radial components constructed at two different linearization energies corresponding to
two different electron bands (or energy windows). We demonstrate that this case can be reduced to
the standard treatment within the LAPW paradigm where the usual basis set is enriched by the basis functions of the
tight binding type, which go to zero with zero derivative at the
sphere boundary.
We show that the task is closely related with the problem of extended core states which is currently solved
by applying the LAPW method with local orbitals (LAPW+LO).
In comparison with LAPW+LO,
the number of supplemented basis functions in our approach is doubled,
which opens up a new channel for the extension of the LAPW and LAPW+LO basis sets.
The appearance of new supplemented basis functions absent in the LAPW+LO treatment
is closely related with the existence of the $\dot{u}_l-$component
in the canonical LAPW method.
We discuss properties of additional tight binding basis
functions and apply the extended basis set for computation of
electron energy bands of lanthanum (face and body centered
structures) and hexagonal close packed lattice of cadmium.
We demonstrate that the new treatment gives lower total
energies in comparison with both canonical LAPW and LAPW+LO, with the energy difference
more pronounced for intermediate and poor LAPW basis sets.
\end{abstract}


\maketitle

\section{INTRODUCTION}

The choice of a basis set which may first appear as ``the black art" \cite{Sza} is being constantly debated within
the quantum chemistry community.
It is well known that in molecular calculations there are two main groups of molecular basis sets
introduced by Pople and collaborators (see \cite{Sza} and references therein) and more recently by Dunning \cite{Dun}.
Both groups supply us with the whole hierarchy of basis sets, where at each step we can enrich the main set
by polarization functions of high orbital type or by diffuse functions.
For example, one has to add polarization functions if polarization effects are expected to be important, or
diffuse functions if we want to refine the description of extended molecular states.
Not surprisingly, the actual choice of a basis set depends on the task to be solved and is considered as a difficult problem.
For heavy and laborous calculations the choice of a basis set is crucial since on one hand we want to obtain
a reliable result and on the other hand minimize the computer time to achieve the goal.

In contrast to this complicated hierarchy of molecular basis sets, the choice of bases in electronic band structure calculations
and here we imply mainly the linear augmented plane wave (LAPW) method \cite{And,Koe,blapw}, seems rather simple.
The number of augmented plane waves is commonly determined by the parameter $R_{MT} K_{max}$, where $R_{MT}$ is the smallest muffin-tin (MT)
radius, and $K_{max}$ is the maximal value of the plane wave vector. The $K_{max}$ value implies that the
kinetic energy cut off is $K_{max}^2/2$ (in atomic units). However, in practice some band structure calculations
can not be carried out without so called local orbitals (LO) \cite{wien2k}.
Such situations occur in systems with  semicore electron states which cannot be fully confined within
the MT-spheres. The problem of extended core states and its relation to our approach is discussed in detail in the next section.
What concerns us here is that the introduction of local orbitals represents an extension of the canonical LAPW basis set
albeit the form of new basis states (local orbitals) is very different from the standard augmented plane wave basis function.
The LAPW+LO scheme proposed by Singh \cite{Sin-2}, \cite{wien2k} is practical, but the way it has been introduced is not
fully satisfactory. The form of the local orbital functions is not derived from a general approach, and arguments
for adding LO basis states are purely variational.

In the present paper we show that the appearance of supplemented basis functions can be understood as a result
of refinement of the LAPW band scheme when in an effort to increase its accuracy we use two linearization energies
(corresponding to two electron bands). We will demonstrate that new basis states are of two types.
The first group gives the local orbitals in the form suggested by Sing \cite{Sin-2}.
however, the basis functions of second type have different form which is not used in the LAPW+LO method.
Therefore, the canonical LAPW basis set and also the LAPW+LO basis set can be extended to a more
complete basis set.
New basis functions and consequences of their introduction are closely examined in the present work.

The paper is organized as follows. We start with revisiting the problem of
extended core states which gives rise to the LAPW+LO scheme and
formulate our initial statement for the refinement of the LAPW method, Sec.\ \ref{semi}.
In Sec.\ \ref{Meth} we present
our method which results in adding tight binding basis
functions to the canonical LAPW basis set.
In Sec.\ \ref{App} we apply the method to electron band
structure calculations of the face centered and body centered
lattice of lanthanum and the hexagonal close packed lattice of
cadmium. Our conclusions are summarized in Sec.\ \ref{Con}.

\section{The problem of extended core states}
\label{semi}

The linear augmented plane wave (LAPW) method \cite{And,Koe,blapw}
is probably the most precise method for electronic band structure
calculations and is widely used for the calculation of materials
properties \cite{wien2k}.

In the LAPW method \cite{And,Koe,blapw, wien2k} space is partitioned in the region inside the nonoverlapping muffin-tin ($MT$) spheres and
the interstitial region $I$.
The basis functions $\phi_n(\vec{k},\,\vec{R})$ where $n=1,2,...,N$ are given by
\begin{equation}
  \phi_n(\vec{k},\,\vec{R}) = \left\{ \begin{array}{ll} V^{-1/2} \, exp(i(\vec{k} + \vec{K}_n) \vec{R}), & \vec{R} \in I   \\
                                                    \sum_{l,m} {\cal R}_{l,m}^{n,\alpha}(r,\, E_l)\, Y_{l,m}(\hat{r}), & \vec{R} \in MT(\alpha)  \end{array} \right.
                                                    \label{i1}
\end{equation}
with radial parts
\begin{equation}
  {\cal R}_{l,m}^{n,\alpha}(r,\, E_l) = A^{n,\alpha}_{l,m}\, u_l(r, E_l) + B^{n,\alpha}_{l,m}\, \dot{u}_l(r, E_l) .
\label{i1a}
\end{equation}
Here the index $\alpha$ refers to the type of atom (or
$MT$-sphere) in the unit cell, the radius $r$ is counted from the
center $\vec{R}_{\alpha}$ of the sphere $\alpha$ (i.e.
$\vec{r}=\vec{R}-\vec{R}_{\alpha}$), $V$ is the volume of the unit
cell. Radial functions $u_{l,m}(r,E_l)$ are solutions of the
Schr\"{o}dinger equation in the spherically averaged crystal
potential computed at the linearization energy $E_l$, and
$\dot{u}_l(r,E_l)$ is the derivative of $u_{l,m}$ with respect to
$E$ at $E_l$. The coefficients $A^n_{l,m}$ and $B^n_{l,m}$ are
found from the condition that the basis function $\phi_n$ is
continuous with continuous derivative at the sphere boundary,
$r=R_{MT}^{\alpha}$ ($R_{MT}^{\alpha}$ is the radius of the
$MT$-sphere $\alpha$). In the following for compactness we omit
the index $\alpha$ and restore it when needed. Linearization
energies $E_l$ are chosen close to average values of corresponding
band energies or to the Fermi level. The extended electron basis
states defined by Eq.\ (\ref{i1}) as a rule are orthogonal to the
core states. This is a consequence of the relation
\begin{eqnarray}
 \int_0^{R_{MT}} {\cal U}_v(r)\, {\cal U}_c(r) r^2 dr = \frac{R_{MT}^2}{2(E_c - E_v )} \nonumber \times\\
  \left( {\cal U}_c(R_{MT}) \frac{\partial {\cal U}_v(R_{MT})}{\partial r} - {\cal U}_v(R_{MT}) \frac{\partial {\cal U}_c(R_{MT})}{\partial r} \right) ,
\label{i2}
\end{eqnarray}
applied to a core state with orbital quantum numbers $l,m$ and the
radial wave function ${\cal U}_c(r)$, and a partial radial
function of valence state, ${\cal U}_v(r)={\cal
R}_{l,m}^{n,\alpha}(r,\, E_l)$, Eq.\ (\ref{i1a}), with the same
angular dependence. Notice that Eq.\ (\ref{i2}) ensures the
orthogonality between the extended and core states if two
conditions at the sphere boundary are satisfied for each of the
core states,
\begin{subequations}
\begin{eqnarray}
  & & {\cal U}_c(R_{MT}) = 0, \label{i3a} \\
  & &  \frac{\partial {\cal U}_c(R_{MT})}{\partial r} = 0 .  \label{i3b}
\end{eqnarray}
\end{subequations}
Although these conditions are met for a great number of cases, they are violated for so called semicore states
that are not fully contained in the muffin-tin sphere \cite{Mat,Sin-1,Sin-2}.

Semicore states leaking out of the $MT$-regions should be treated
as extended states. This in turn requires that the linearization
energy $E_l$ is chosen near the energy of the semicore level, $E_l
\approx E_c$, because the LAPW basis describes only states near
$E_l$ well. However, as $E_c$ is quite far from the Fermi energy
$E_F$ and the valence band energy, the choice $E_l \approx E_c$
inevitably gives poor description for partial $l=l_c$ valence
states. On the other hand, the option $E_l =
E_v$ is not satisfactory for the semicore states situated
substantially deeper in energy. As discussed in Refs.\
\cite{Sin-1,Sin-2,Bla-1,Sin-3,Sin-4,Goe} there is no simple
solution to this dilemma. Even worse, in many cases the attempt to
use a single value of $E_l$ for both valence and semicore states
leads to the appearance of so called ``ghost bands"
\cite{Sin-1,Sin-2} giving false band energy positions. As a remedy
one can divide the energy spectrum in two windows (energy panels)
and use two different sets of $E_l$ for calculations of semicore
and valence states, respectively \cite{blapw}. This technique
however is also not fully satisfactory because now there is no
single Hamiltonian matrix for the problem and strict orthogonality
between electron states belonging to different energy windows is
not guaranteed. Ideally, in the MT-region there should be two
different types of radial components with the same angular
dependence $l=l_c$. That is, in Eq.\ (\ref{i1})
\begin{subequations}
\begin{equation}
 {\cal R}_{l,m}^{n}(r) = {\cal R}_{l,m}^{(1),n}(r) + {\cal R}_{l,m}^{(2),n}(r) ,
\label{i4}
\end{equation}
where
\begin{equation}
  {\cal R}_{l,m}^{(1),n}(r,\, E_l^{(1)}) = A^{(1),n}_{l,m}\, u_l^{(1)}(r, E_l^{(1)}) + B^{(1),n}_{l,m}\, \dot{u}_l^{(1)}(r, E_l^{(1)})
\label{i4a}
\end{equation}
refers to the semicore states with $E_l^{(1)} = E_c$, and
\begin{equation}
  {\cal R}_{l,m}^{(2),n}(r,\, E_l^{(2)}) = A^{(2),n}_{l,m}\, u_l^{(2)}(r, E_l^{(2)}) + B^{(2),n}_{l,m}\, \dot{u}_l^{(2)}(r, E_l^{(2)})
\label{i4b}
\end{equation}
refers to the valence states with $E_l^{(2)} = E_v \approx E_F$.
Both states, Eqs.\ (\ref{i4a}) and (\ref{i4b}), should merge to a single
$l_c-$wave component of the
plane wave
\begin{equation}
  \phi_n(\vec{k},\,\vec{R}) = V^{-1/2} \, exp(i(\vec{k} + \vec{K}_n) \vec{R})
\label{i4c}
\end{equation}
\end{subequations}
at the surface of MT sphere.
Now, however the boundary problem becomes ill-defined, because for the $l=l_c$ component
there are four coefficients, $A^{(1),n}_{l,m}$, $B^{(1),n}_{l,m}$, $A^{(2),n}_{l,m}$ and $B^{(2),n}_{l,m}$
for only two boundary conditions.

In Ref.\ \cite{Sin-2} Singh has proposed to increase the number of boundary conditions to four
by matching the value of the basis function and its first three radial derivatives
at the sphere surface.
This gives rise to super-linearized APW method denoted as SLAPW-4 \cite{Sin-2}
because four functions, four coefficients and four boundary conditions are involved.
In a simpler super-linearized modification called SLAPW-3 \cite{Sin-2} the first radial part ${\cal R}_{l,m}^{(1),n}$, Eq.\ (\ref{i4a}),
is supplemented by only one function $u_l^{(2)}$
(instead of ${\cal R}_{l,m}^{(2),n}$, Eq.\ (\ref{i4b})).
The three coefficients ($A^{(1),n}_{l,m}$, $B^{(1),n}_{l,m}$, $A^{(2),n}_{l,m}$) are determined by requiring continuity of the basis function and its two derivatives.

In comparison with standard LAPW method in both SLAPW
modifications there are additional requirements for the plane wave
convergence. Indeed, the plane wave expansion of the interstitial
region must converge either to the correct second and third
derivative (SLAPW-4) or to the second derivative (SLAPW-3).
Because of that more plane waves are needed to satisfy these
additional requirements, the plane wave energy cutoff parameter
should be increased and calculations become much more costly
\cite{Sin-2},\cite{blapw}.

To circumvent the problem and improve the LAPW efficiency Singh
put forward a third approach based on local orbitals (LAPW+LO)
\cite{Sin-2}. In the LAPW+LO approach the same three radial
functions as in SLAPW-3 are used (i.e. $u_l^{(1)}$,
$\dot{u}_l^{(1)}$ and $u_l^{(2)}$), but the coefficient of
$u_l^{(2)}$ is fixed (say, $A^{(2),n}_{l,m}=1$) and the two
remaining coefficients ($A^{(1),n}_{l,m}$, $B^{(1),n}_{l,m}$) are
found from the conditions that the local orbital goes to zero with
zero derivative at the sphere boundary. Nowadays, LAPW+LO is
widely used for band structure calculations of solids with
semicore states \cite{blapw,wien2k}. However, conceptually the
LAPW+LO method is understood as a procedure giving additional
variational freedom through an increase of the number of basis
functions. It is not clear why additional basis functions should
include these particular components (i.e. $u_l^{(1)}$,
$\dot{u}_l^{(1)}$ and $u_l^{(2)}$). The proposed zero boundary conditions
for local functions are not derived from a general physical statement.

Inspired by the LAPW+LO method \cite{Sin-2} in the present study
we formulate a more general approach to the problem. Unlike the
LAPW+LO approach which uses variational arguments for its
foundation, we will derive supplemented basis states from the
initial requirement that two different radial functions (i.e.
${\cal R}_{l,m}^{(1),n}$ and ${\cal R}_{l,m}^{(2),n}$, Eqs.\
(\ref{i4a}), (\ref{i4b})), having the same angular part merge in a
single plane wave function $\phi_n$, Eq.\ (\ref{i4c}), in the
interstitial region. Unlike SLAPW-4 or SLAPW-3 we retain only two
joining conditions across the $MT$-sphere boundary. As a result,
we will obtain two types of supplementary tight-binding basis
functions (see Eqs.\ (\ref{m10b}) and (\ref{m10c}) below),
satisfying Bloch's theorem.

\section{Description of the method}
\label{Meth}

As discussed in Sec.\ \ref{semi}, in the case of semicore
states we have two types of radial solutions in the MT-region with
the same angular dependence $Y_{l,m}(\hat{r})$ but different linearization energies $E_l^{(1)}$
and $E_l^{(2)}$:
${\cal R}_{l,m}^{(1),n}(r,\, E_l^{(1)})$, Eq.\ (\ref{i4a}), and ${\cal
R}_{l,m}^{(2),n}(r,\, E_l^{(2)})$, Eq.\ (\ref{i4b}). One of the radial
functions can refer to extended states, i.e. $R_e(r)={\cal
R}_{l,m}^{(1),n}(r,\, E_l^{(1)})$, while the other can refer to supplementary
angular states $R_s(r)={\cal R}_{l,m}^{(2),n}(r,\, E_l^{(2)})$. As we will see
later in Sec.\ \ref{App} in practice we describe the semicore
states as extended states with $E_l = E_{core}$ and valence states
with the same $l$ as supplementary
states for which $E_l=E_v$.
(For metals one can take $E_l=E_v \approx E_F$.)
Since in the interstitial
$I$-region both types of solutions are represented by the plane
wave function $\phi_n(\vec{k},\,\vec{R})$, Eq.\ (\ref{i4c}), they
become indistinguishable there. In the LAPW method there are two
matching conditions (for the function and its derivative) on the
sphere boundary. Therefore, in our case we have
\begin{subequations}
\begin{eqnarray}
   A_e u_e + B_e \dot{u}_e + A_s u_s + B_s \dot{u}_s =   \nonumber \\
   \frac{4\pi}{\sqrt{V}} i^l j_l(k_n R_{MT}) Y_{l,m}^*(\hat{k}_n)\, e^{i \vec{k}_n \vec{R}_{\alpha}}, \label{m1a} \\
  A_e u'_e + B_e \dot{u}'_e + A_s u'_s + B_s \dot{u}'_s = \nonumber \\
  \frac{4\pi}{\sqrt{V}} i^l j'_l(k_n R_{MT}) Y_{l,m}^*(\hat{k}_n)\, e^{i \vec{k}_n \vec{R}_{\alpha}} .  \label{m1b}
\end{eqnarray}
\end{subequations}
Here we adopt short notations $A_e = A^{(1),n}_{l,m}$, $A_s = A^{(2),n}_{l,m}$, $u_e = u_l^{(1)}(R_{MT},E_l^{(1)})$, $u_s = u_l^{(2)}(R_{MT},E_l^{(2)})$,
$u'_e = \partial u_l^{(1)}(R_{MT},E_l^{(1)})/\partial r$, $u'_s = \partial u_l^{(2)}(R_{MT},E_l^{(2)})/ \partial r$,
and have used the Rayleigh expansion of the plane wave $\phi_n$ on the sphere surface.
Since there are four coefficients ($A_e$, $B_e$, $A_s$ and $B_s$) and only two equations, it is clear that the general solution to Eqs.\ (\ref{m1a}), (\ref{m1b}),
forms a two dimensional linear space with two linear independent basis vectors.

Further, introducing standard LAPW quantities $a_e = a^{(1),n}_l$, $b_e = b^{(1),n}_l$, where
\begin{subequations}
\begin{eqnarray}
   A_e =  \frac{4\pi}{\sqrt{V}} i^l R_{MT}^2\, Y_{l,m}^*(\hat{k}_n)\, e^{i \vec{k}_n \vec{R}_{\alpha}}\, a_e , \label{m2a} \\
   B_e =  \frac{4\pi}{\sqrt{V}} i^l R_{MT}^2\, Y_{l,m}^*(\hat{k}_n)\, e^{i \vec{k}_n \vec{R}_{\alpha}}\, b_e .  \label{m2b}
\end{eqnarray}
\end{subequations}
and analogous relations for $a_s$, $b_s$, we rewrite Eqs.\ (\ref{m1a}), (\ref{m1b}) as
\begin{subequations}
\begin{eqnarray}
   a_e u_e + b_e \dot{u}_e + a_s u_s + b_s \dot{u}_s = j_l(k_n R_{MT}) \frac{1}{R_{MT}^2} , \label{m3a} \\
  a_e u'_e + b_e \dot{u}'_e + a_s u'_s + b_s \dot{u}'_s =  j'_l(k_n R_{MT}) \frac{1}{R_{MT}^2} .  \label{m3b}
\end{eqnarray}
\end{subequations}

Notice that the standard LAPW solution for $a_e = a_e^0$ and $b_e = b_e^0$ without supplementary states,
i.e. when $a_s=0$, $b_s=0$, can be found from
the following system
\begin{subequations}
\begin{eqnarray}
   a_e^0 u_e + b_e^0 \dot{u}_e = j_l(k_n R_{MT}) \frac{1}{R_{MT}^2} , \label{m4a} \\
  a_e^0 u'_e + b_e^0 \dot{u}'_e =  j'_l(k_n R_{MT}) \frac{1}{R_{MT}^2} .  \label{m4b}
\end{eqnarray}
\end{subequations}
Defining auxiliary quantities $t_a$ and $t_b$
\begin{subequations}
\begin{eqnarray}
  t_a =  a_e - a_e^0 , \label{m5a} \\
  t_b =  b_e - b_e^0  .  \label{m5b}
\end{eqnarray}
\end{subequations}
and subtracting Eq.\ (\ref{m4a}) from Eq.\ (\ref{m3a}), and  Eq.\ (\ref{m4b}) from Eq.\ (\ref{m3b})
we arrive at
\begin{subequations}
\begin{eqnarray}
   t_a u_e + t_b \dot{u}_e + a_s u_s + b_s \dot{u}_s = 0 , \label{m6a} \\
   t_a u'_e + t_b \dot{u}'_e + a_s u'_s + b_s \dot{u}'_s =  0 .  \label{m6b}
\end{eqnarray}
\end{subequations}

The solution to Eqs.\ (\ref{m6a}), (\ref{m6b}) can be found from
the following two systems,
\begin{subequations}
\begin{eqnarray}
   \left\{ \begin{array}{l} a_{s,1} u_s + b_{s,1} \dot{u}_s = -u_e \\
                             a_{s,1} u'_s + b_{s,1} \dot{u}'_s = -u'_e  \end{array} \right. ,
                                                    \label{m7a} \\
   \left\{ \begin{array}{l} a_{s,2} u_s + b_{s,2} \dot{u}_s = -\dot{u}_e \\
                             a_{s,2} u'_s + b_{s,2} \dot{u}'_s = -\dot{u}'_e  \end{array} \right. .
                                                    \label{m7b}
\end{eqnarray}
\end{subequations}
Solutions to the systems (\ref{m7a}) and (\ref{m7b}) are quoted  explicitly in Appendix~\ref{appA}, Eqs.\ (\ref{a1a})-(\ref{a2b}).
Having found $a_{s,i}$ and $b_{s,i}$ ($i=1,2$), we write the general solution to Eqs.\ (\ref{m3a}), (\ref{m3b})
as
\begin{subequations}
\begin{eqnarray}
 & & a_e = a_e^0 + t_a , \label{m8a} \\
 & & b_e = b_e^0 + t_b ,  \label{m8b} \\
 & & a_s = t_a \, a_{s,1} + t_b \, a_{s,2} , \label{m8c} \\
 & & b_s = t_a \, b_{s,1} + t_b \, b_{s,2} , \label{m8d}
\end{eqnarray}
\end{subequations}
where $t_a$ and $t_b$ are arbitrary numbers.
The full radial component ${\cal R}_{l,m}^{n}(r)$ of the basis function inside the $MT$-sphere $\alpha$, Eq.\ (\ref{i4}), is written as
\begin{eqnarray}
   {\cal R}_{l,m}^{n}(r) \sim a_e^0\, u_e + b_e^0\, \dot{u}_e + t_a \, (u_e + a_{s,1}\, u_s + b_{s,1}\, \dot{u}_s) \nonumber \\
    + t_b \, (\dot{u}_e + a_{s,2}\, u_s + b_{s,2}\, \dot{u}_s) . \quad \label{m9}
\end{eqnarray}
(Here notations $u_e=u_e(r)$, $u_s=u_s(r)$ etc. refer to radial
functions.)

Notice that since the coefficients $t_a$ and $t_b$ are arbitrary, they should be found from the standard variational procedure by requiring
the minimization of the LAPW ground state energy.
Furthermore, the form (\ref{m9}) suggests considering three linear independent radial parts (i.e. $R_{l,m}^{e}(r)$, $R_{l,m}^{s,1}(r)$, $R_{l,m}^{s,2}(r)$)
instead of the single function ${\cal R}_{l,m}^{n}(r)=R_{l,m}^{e}(r) + t_a\, R_{l,m}^{s,1}(r) + t_b\, R_{l,m}^{s,2}(r)$.
Explicitly,
\begin{subequations}
\begin{eqnarray}
   & & R_{l,m}^{e}(r) = C_n\, e^{i \vec{k}_n \vec{R}_{\alpha}} ( a_e^0\, u_e + b_e^0\, \dot{u}_e ),   \label{m10a} \\
   & & R_{l,m}^{s,1}(r) = C_n\, e^{i \vec{k}_n \vec{R}_{\alpha}} ( u_e + a_{s,1}\, u_s + b_{s,1}\, \dot{u}_s ),  \label{m10b} \\
   & & R_{l,m}^{s,2}(r) = C_n\, e^{i \vec{k}_n \vec{R}_{\alpha}} ( \dot{u}_e + a_{s,2}\, u_s + b_{s,2}\, \dot{u}_s ). \quad \quad \label{m10c}
\end{eqnarray}
\end{subequations}
Here $u_e=u_e(r)$, $u_s=u_s(r)$ etc. are corresponding radial functions and
\begin{eqnarray}
    C_n = \frac{4\pi}{\sqrt{V}} i^l R_{MT}^2\, Y_{l,m}^*(\hat{k}_n) . \label{i14}
\end{eqnarray}
The first function, Eq.\ (\ref{m10a}), is in fact the standard radial part of the $l-$type, $R_{l,m}^{e}={\cal R}_{l,m}^{n,\alpha}$, Eq.\ (\ref{i1a}),
entering the usual LAPW basis function $\phi_n(\vec{k},\,\vec{R})$, Eq.\ (\ref{i1}).
Its coefficients $a_e^0$ and $b_e^0$ are given by the LAPW boundary relations, Eqs.\ (\ref{m4a}), (\ref{m4b}),
Two other functions however are very different from $\phi_n(\vec{k},\,\vec{R})$ and should be included to the LAPW
basis set as extra basis states,
\begin{eqnarray}
  \phi_{s,i}(\vec{k},\,\vec{R}) = Y_l^m(\hat{r}) \, R_{l,m}^{s,i}(r) , \label{m11}
\end{eqnarray}
where $i=1,2$.
The important thing is that their coefficients $a_{s,1}$, $b_{s,1}$ are
found from Eq.\ (\ref{m7a}), while coefficients $a_{s,2}$, $b_{s,2}$ from Eq.\ (\ref{m7b}).
In respect to two new functions $R_{l,m}^{s,i}$, Eqs.\ (\ref{m7a}), (\ref{m7b}) impose the following boundary conditions
\begin{subequations}
\begin{eqnarray}
 R_{l,m}^{s,i}(r) = 0 , \label{m12a}  \\
 \frac{\partial R_{l,m}^{s,i}(r)}{\partial r} = 0 . \label{m12b}
\end{eqnarray}
\end{subequations}
These relations have a simple interpretation: new supplementary basis functions $\phi_{s,i}(\vec{k},\,\vec{R})$
are required to be orthogonal to the standard LAPW radial functions.
We want to stress here, that the conditions (\ref{m12a}) and (\ref{m12b}) are not assumed or introduced at our will.
They are derived from the initial equations (\ref{m1a}), (\ref{m1b}) [or equivalently from Eqs.\ (\ref{m3a}), (\ref{m3b})]
and are used to obtain the general solution, Eqs.\ (\ref{m8a})-(\ref{m8d}).

From Eq.\ (\ref{m10b}), (\ref{m10c}) and (\ref{i14}) it follows that the supplementary basis states
in principle depend on the index $n$, i.e. $\phi_{s,i}(\vec{k},\,\vec{R}) = \phi_{s,i}(\vec{k}_n,\,\vec{R})$.
(We recall that $\vec{k}_n=\vec{k}+\vec{K}_n$, where $\vec{K}_n$ is a vector of the reciprocal lattice.)
However, since all functions $\phi_{s,i}(\vec{k}_n,\,\vec{R})$ with different index $n$ have the same radial part,
${\cal U}^{s,1}(r) = u_e + a_{s,1}\, u_s + b_{s,1}\, \dot{u}_s$ for $i=1$, or
${\cal U}^{s,2}(r) = \dot{u}_e + a_{s,2}\, u_s + b_{s,2}\, \dot{u}_s $ for $i=2$,
they are simply proportional to each other, $\phi_{s,i}(\vec{k}_{n},\,\vec{R}) \sim \phi_{s,i}(\vec{k}_{n'},\,\vec{R})$.
Therefore, to avoid the linear dependence we should choose only one set of functions $\phi_{s,i}(\vec{k}_n,\,\vec{R})$ corresponding to a single index $n$.
The obvious choice is to use the function $\phi_{s,i}(\vec{k},\,\vec{R})$ with $n=0$, $\vec{k}_0 = \vec{k}$ and
the reciprocal lattice vector $\vec{K}_0 = 0$.
In that case the coefficient $C_0 = \pi/V\, i^l R_{MT}^2\, Y_{l,m}^*(\hat{k})$ [compare with Eq.\ (\ref{i14})]
can be further rationalized by omitting the multiplier $Y_{l,m}^*(\hat{k})$ [or equivalently, including it in factors $t_a$ and $t_b$,
Eq.\ (\ref{m5a}), (\ref{m5b})]. Thus, we substitute $C_0$ with
\begin{eqnarray}
 {\cal C}_0 = \frac{4\pi}{\sqrt{V}} i^l R_{MT}^2\ . \label{i17}
\end{eqnarray}
(In principle, since the local function is not orthonormal, we can simply put ${\cal C}_0 = 1$, but the form (\ref{i17}) being similar to
the constant coefficient for the standard LAPW basis function, simplifies some expressions for programming.)

To study the transformational properties of supplementary basis functions $\phi_{s,i}(\vec{k},\,\vec{R})$, Eq.\ (\ref{m11}),
we first rewrite them in the following form,
\begin{eqnarray}
  \phi_{s,i}(\vec{k},\,\vec{R}) =  e^{i \vec{k} \vec{R}_{\alpha}} \, \psi_i^{l,m}(\vec{R}-\vec{R}_{\alpha}) , \label{m12'}
\end{eqnarray}
where for each site $\alpha$ we have introduced two local functions ($i=1,2$) of the $l,m$-type,
\begin{eqnarray}
  \psi_i^{l,m}(\vec{R}-\vec{R}_{\alpha})= {\cal C}_0 \, {\cal U}^{s,i}(r) \, Y_l^m(\hat{r}) . \label{m14}
\end{eqnarray}
Notice that each local function $\psi_i^{l,m}(\vec{R}-\vec{R}_{\alpha})$
is strictly confined inside the $MT$-sphere $\alpha$, because both ${\cal U}^{s,i}(r)$ and $R_l^{s,i}(r)$
satisfy the boundary conditions (\ref{m12a}), (\ref{m12b}).
We can then extend the function $\phi_{s,i}(\vec{k},\,\vec{R})$ to the interstitial region ($\vec{R} \in I$) by requiring $\phi_{s,i}(\vec{k},\,\vec{R}) = 0$.
For the whole crystal we thus have
\begin{eqnarray}
    \begin{array}{ll} \phi_{s,i}(\vec{k},\,\vec{R}) = \sum_{\alpha}  e^{i \vec{k} \vec{R}_{\alpha}} \, \psi_i^{l,m}(\vec{R}-\vec{R}_{\alpha}),
                                     & \vec{R} \in MT,   \\
                              \phi_{s,i}(\vec{k},\,\vec{R}) = 0, &    \vec{R} \in I.
                                                      \end{array}         \label{m16}
\end{eqnarray}
This is a clear manifestation of the tight binding wave function.
The multiplier $e^{i \vec{k} \vec{R}_{\alpha}}$
in Eqs.\ (\ref{m16}) and (\ref{m12'}) ensures that the supplementary wave functions $\phi_{s,i}(\vec{k},\,\vec{R})$
obey the Bloch theorem.
It is worth noting that usually the tight-binding description is spoiled by the presence of overlap
between tails of wave function centered at neighboring sites. In the present method the tight-binding functions,
Eqs.\ (\ref{m16}) and (\ref{m12'}), are free from this disadvantage because the local functions
(and their first derivatives) go to zero at the sphere boundary and the overlap is absent.
Thus, supplementary tight-binding functions can be considered as additional basis states orthogonal
to the standard LAPW basis set.

All matrix elements between the supplementary basis functions $\phi_{s,i}$ in the spherically symmetric potential
are quoted explicitly in Appendix~\ref{appB},
and all matrix elements between $\phi_{s,i}$ and standard LAPW basis functions $\phi_n$ are listed in Appendix~\ref{appC}.
For briefness we do not quote here the partial charges and electron density associated with supplementary basis states.
They are tightly connected with the overlap matrix elements given by Eqs.\ (\ref{b1a}), (\ref{b3a}) and (\ref{b4a})
of Appendix~\ref{appB}, and Eqs.\ (\ref{c2a}), (\ref{c3a}) of Appendix~\ref{appC}.
Concerning the full potential expressions for the extended basis set it is worth noting that after some algebraic transformations the equations can be obtained
by selecting in standard FLAPW equations the contributions with the orbital indices $l,m$ referring to the components of supplemented states
and combining them together according to Eqs.\ (\ref{m10b}), (\ref{m10c}), (\ref{m11}).

The tight binding basis functions have a very important and practical property: they work even in the case when their expansion energy $E_s$
lies not far from the LAPW linear expansion energy $E_e$.
(We recall that the whole procedure is designed to treat the complicated case
of semicore states when $E_s$ is supposed to be separated from $E_e$ by at least 10 eV.) The limiting case $E_s \approx E_e$ is considered in detail
in Appendix \ref{appD}, and also discussed in calculations of Cd in Sec.\ \ref{app-Cd}.

\section{Practical Implementation}
\label{App}

\subsection{Face centered cubic structure of La}
\label{App-A}

We have applied the method developed in Sec.\ \ref{Meth} to full potential electron band structure calculations of face centered cubic (fcc)
structure of lanthanum.
Atomic lanthanum has completely filled 5$p$ semicore electron shell lying at -22.12 eV which can slightly mix with valence states ($5d$, $6s$, $4f$)
at energies from -3 to -2 eV. For lanthanum here and below we use the Perdew-Burke-Ernzerhof (PBE) \cite {PBE} variant of the generalized gradient approximation (GGA) which gives rather accurate lattice constants for our systems.

We have employed our original version of LAPW code with the potential of general form \cite{Nik}.
Integration in the irreducible part of the Brillouin zone has been performed over 240 special points.
Angular expansions for the electron density and wave function inside MT-sphere have been done up to $L_{max}=8$.
The number of basis functions has been limited by the condition $K_{max} R_{MT} = 9.0$, resulting in 65 basis states.
In addition to the standard LAPW basis functions we have considered 6 supplementary tight binding basis functions with the $p-$angular
dependence, which are located strictly inside the MT-sphere, Sec.\ \ref{Meth}.

\begin{figure}
\resizebox{0.4\textwidth}{!}
{\includegraphics{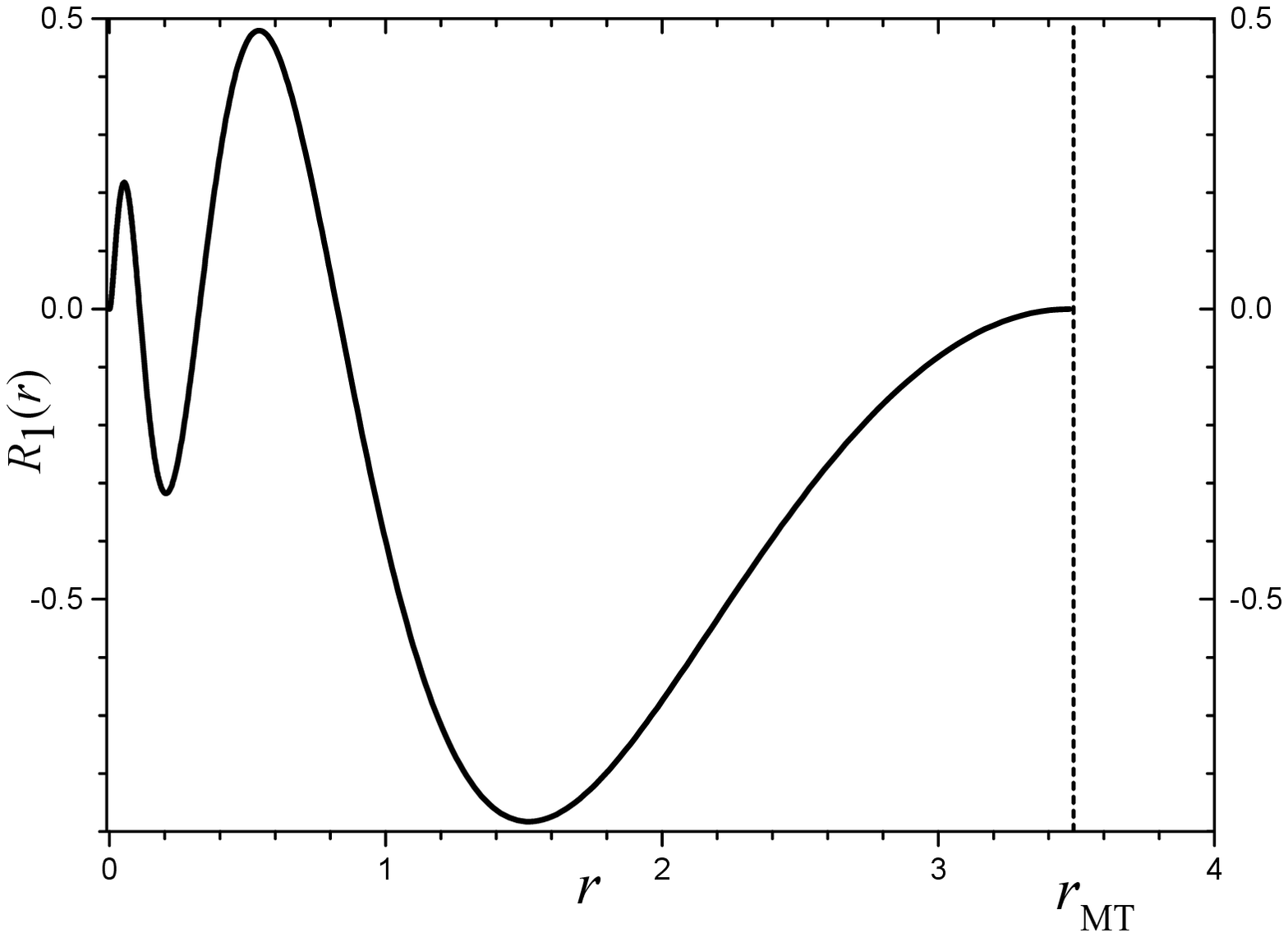}\label{fig1a}}
\resizebox{0.4\textwidth}{!}
{\includegraphics{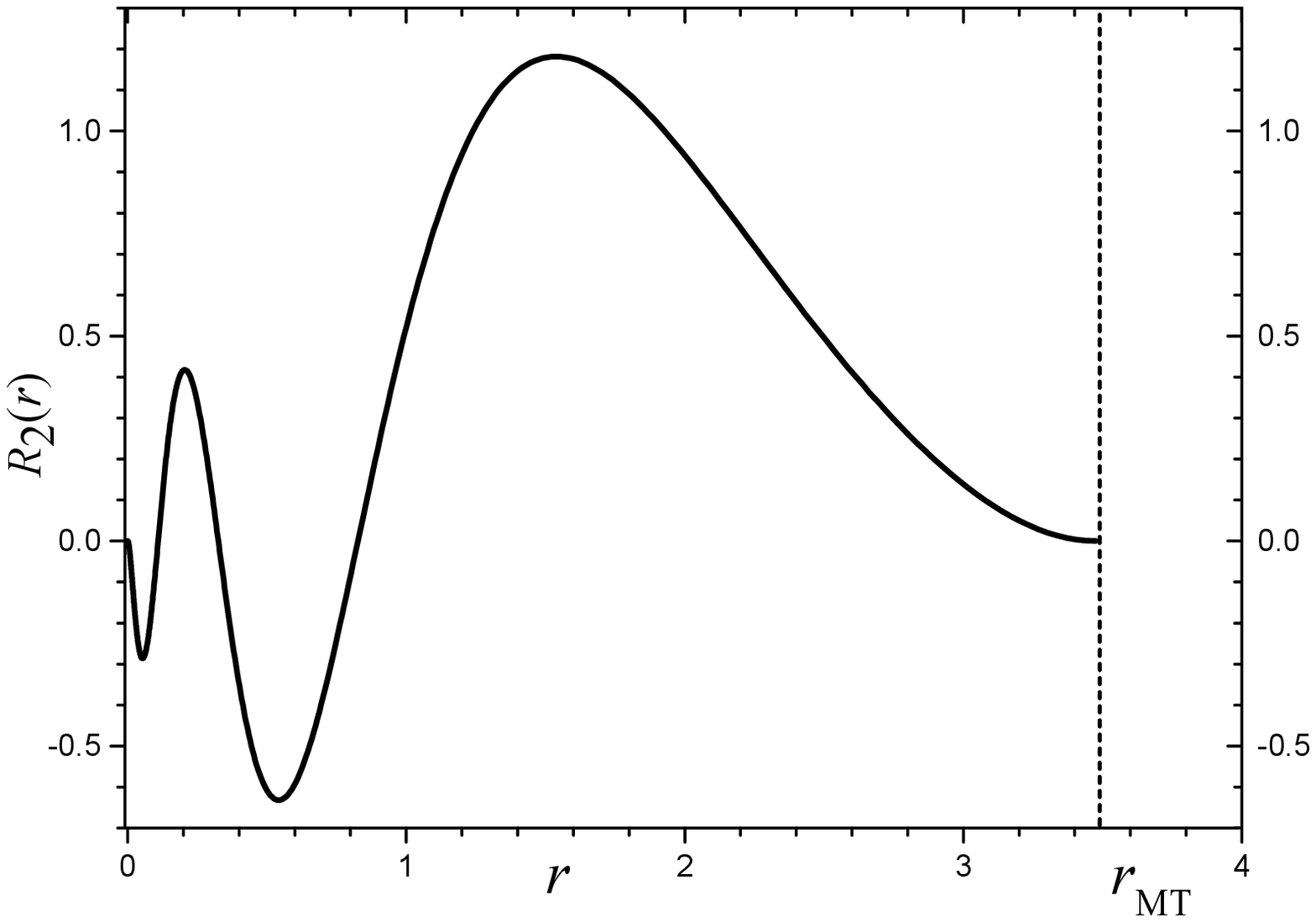}\label{fig1b}}

\caption{
Two supplemented tight binding radial functions of $p-$type for face centered cubic (fcc) structure of lanthanum ($a=5.315$ {\AA}). Radius is given
in {\AA}, $r_{MT}=3.474$~{\AA} stands for the muffin-tin radius.
}
\label{fig1}
\end{figure}

The $5p$ semicore states have been treated as band states with the LAPW linear expansion energy $E_{e}$ lying 0.5 eV
above the $p-$band bottom energy (which is $-9.5$ eV for the equilibrium lattice constant $a=5.315$~{\AA}).
The linear expansion energy for supplementary tight binding $p-$states
has been fixed at 1 eV below the Fermi energy, $E_s(p)=6.10$~eV.
Under these conditions two supplementary radial functions shown in Fig.\ \ref{fig1} are given by
\begin{subequations}
\begin{eqnarray}
  R_1(r)=u_e(r)+a_{s,1} u_s(r) + b_{s,1} \dot{u}_s(r) , \label{p1}
\end{eqnarray}
where $a_{s,1}=0.1940$, $b_{s,1}=0.4041$, and
\begin{eqnarray}
  R_2(r)=\dot{u}_e(r)+a_{s,2} u_s(r) + b_{s,2} \dot{u}_s(r) , \label{p2}
\end{eqnarray}
\end{subequations}
where $a_{s,2}=-1.9609$, $b_{s,2}=1.0705$.
Notice that the number of nodes for both radial functions is three (excluding points with $r=0$ and $r=R_{MT}$) which allows us
to consider these functions as ``compressed'' $5p$ basis states (i.e. with the principal quantum number $n=5$) strictly confined within MT-sphere.
The iteration procedure with supplemented tight binding functions has been stable converging to a self consistent solution without
additional difficulties. Our PBE-GGA calculations result in equilibrium fcc lattice constant $a=5.315$~{\AA}
which compares well with the experimental value, $a_{exp}=5.304$~{\AA}, Ref.\ \onlinecite{fcc-La}.

%
\begin{table}
\caption{ Total energy ($E_{tot}$, in eV) for various basis sets for fcc calculations of La. $a=5.315$ {\AA},
$R_{MT}=3.474$~{\AA}, $E_0=-231170$ eV, $\triangle E= E_{tot}($FLAPW++$)\,-E_{tot}($FLAPW+).
FLAPW++ stands for the present scheme (FLAPW + 2STBFs) with two radial functions
and FLAPW+ for the FLAPW + LO method with a single radial function.
\label{tab1f} }

\begin{ruledtabular}
\begin{tabular}{c | c  c  c  }

$R_{MT} \cdot K_{max}$ & FLAPW++         & FLAPW+       &  $\triangle E$ \\
\tableline
   7.0                 & $E_0-8.6878$    & $E_0-8.5432$ & -0.1446  \\
   7.5                 & $E_0-9.2282$    & $E_0-9.1730$ & -0.0552  \\
   8.0                 & $E_0-9.5362$    & $E_0-9.5301$ & -0.0061  \\
   9.0                 & $E_0-9.5635$    & $E_0-9.5599$ & -0.0036 \\

\end{tabular}
\end{ruledtabular}
\end{table}
%
\begin{table}
\caption{ Energy parameters (in eV) for various basis sets for fcc calculations of La ($a=5.315$ {\AA},
$R_{MT}=3.474$~{\AA}).
FLAPW++ stands for the present scheme (FLAPW + 2STBFs) with two radial functions
and FLAPW+ for the FLAPW + LO method with a single radial function.
\label{tab2f} }

\begin{ruledtabular}
\begin{tabular}{c c | c  c |  c  c }

          &           & \multicolumn{2}{c|}{semicore}    & \multicolumn{2}{c}{valence} \\
          &           & \multicolumn{2}{c|}{$5p-$band}   & \multicolumn{2}{c}{$(spd)-$band} \\
  &  $R_{MT} K_{max}$ &   $E_{bot}$  &  $E_{top}$   & $E_{bot}$   &  $E_F$ \\
\tableline
           & 7.0      & -11.1905 & -9.9254 & 2.7704 & 6.2047 \\
  FLAPW++  & 7.5      & -10.9033 & -9.5956 & 2.9965 & 6.3020 \\
           & 8.0      & -10.0729 & -8.7075 & 3.7651 & 7.0381 \\
           & 9.0      & -9.9972 & -8.6297 & 3.8310 & 7.1009 \\
\tableline
           & 7.0      & -10.9433 & -9.6852 & 2.9275 & 6.3934 \\
  FLAPW+   & 7.5      & -10.7326 & -9.4289 & 3.1170 & 6.4386 \\
           & 8.0      & -9.9564 & -8.5890 & 3.8695 & 7.1478 \\
           & 9.0      & -9.9240 & -8.5550 & 3.8958 & 7.1695 \\

\end{tabular}
\end{ruledtabular}
\end{table}
%
\begin{table}
\caption{ Energy band spectrum (in eV) of fcc La ($a=5.315$ {\AA},
$R_{MT}=3.474$~{\AA}, $R_{MT} \cdot K_{max}=9$) at the $\Gamma$-point
of the Brillouin zone.
FLAPW++ stands for the present scheme (FLAPW + 2STBFs) with two radial functions
and FLAPW+ for the FLAPW + LO method with a single radial function.
\label{tab3f} }

\begin{ruledtabular}
\begin{tabular}{c  c  c  c  }

band & deg. & FLAPW++ & FLAPW+ \\
\tableline
 1    & (3) & -8.6297 & -8.5550 \\
 2    &     &  3.8310 &  3.8958 \\
$E_F$ &     &  7.1009 &  7.1695 \\
 3    &     &  8.5444 &  8.6258 \\
 5    & (3) &  8.6128 &  8.6803 \\

\end{tabular}
\end{ruledtabular}
\end{table}

To compare our treatment with the standard (LAPW+LO) method which uses only the first local function $R_1(r)$, Eq.\ (\ref{p1}),
we have performed a series of calculations, the results of which are summarized in Tables \ref{tab1f}, \ref{tab2f}, \ref{tab3f}.
In all cases the present method gives lower values of the total energy, Table~\ref{tab1f}. However, the energy difference
which amounts to 0.145 eV for the poor basis set ($K_{max} R_{MT} = 7$) becomes smaller for the intermediate basis sets
and decreases to a small value (0.004 eV) for the best basis set ($K_{max} R_{MT} = 9$).
Nevertheless, even this difference
is clearly noticeable in energy band characteristics. In particular, band energy spectrum
demonstrates that the energy difference is of $\approx$0.07 eV for the best basis set, Tables \ref{tab2f}, \ref{tab3f}.
It is worth noting that for all basis sets
the present method results gives smaller band energies, Table \ref{tab2f}.

\subsection{Body centered cubic structure of La}

For the
body centered cubic (bcc) phase of lanthanum we have used a plane wave cut off parameter $K_{max} R_{MT} = 9$
(79 basis states plus 6 additional tight binding
$p-$states), $L_{max}=8$ for the expansion of the electron density
and wave functions inside MT-spheres, 285 special points
in the irreducible part of the Brillouin zone during the
self-consistent procedure.

As for fcc-La the LAPW linear expansion energy for the extended
$p-$states of bcc-La has been has been chosen at 0.5 eV
above the $p-$band bottom energy (i.e. $E_{e}(p)=-9.29$~eV for the equilibrium lattice constant $a=4.243$ {\AA}).
The linear expansion energy for
the supplemented tight binding $p-$states has been fixed at 1.0 eV below the Fermi
energy ($E_s(p)=6.32$~eV). Two supplemented radial functions, quoted in Eqs.\ (\ref{p1}) and (\ref{p2}),
are defined by the coefficients $a_{s,1}=0.1502$, $b_{s,1}=0.4844$ for $R_1(r)$,
and $a_{s,2}=-1.8251$, $b_{s,2}=0.7718$ for $R_2(r)$.

\begin{figure}[t]
\subfloat[]{\includegraphics[width=\linewidth]{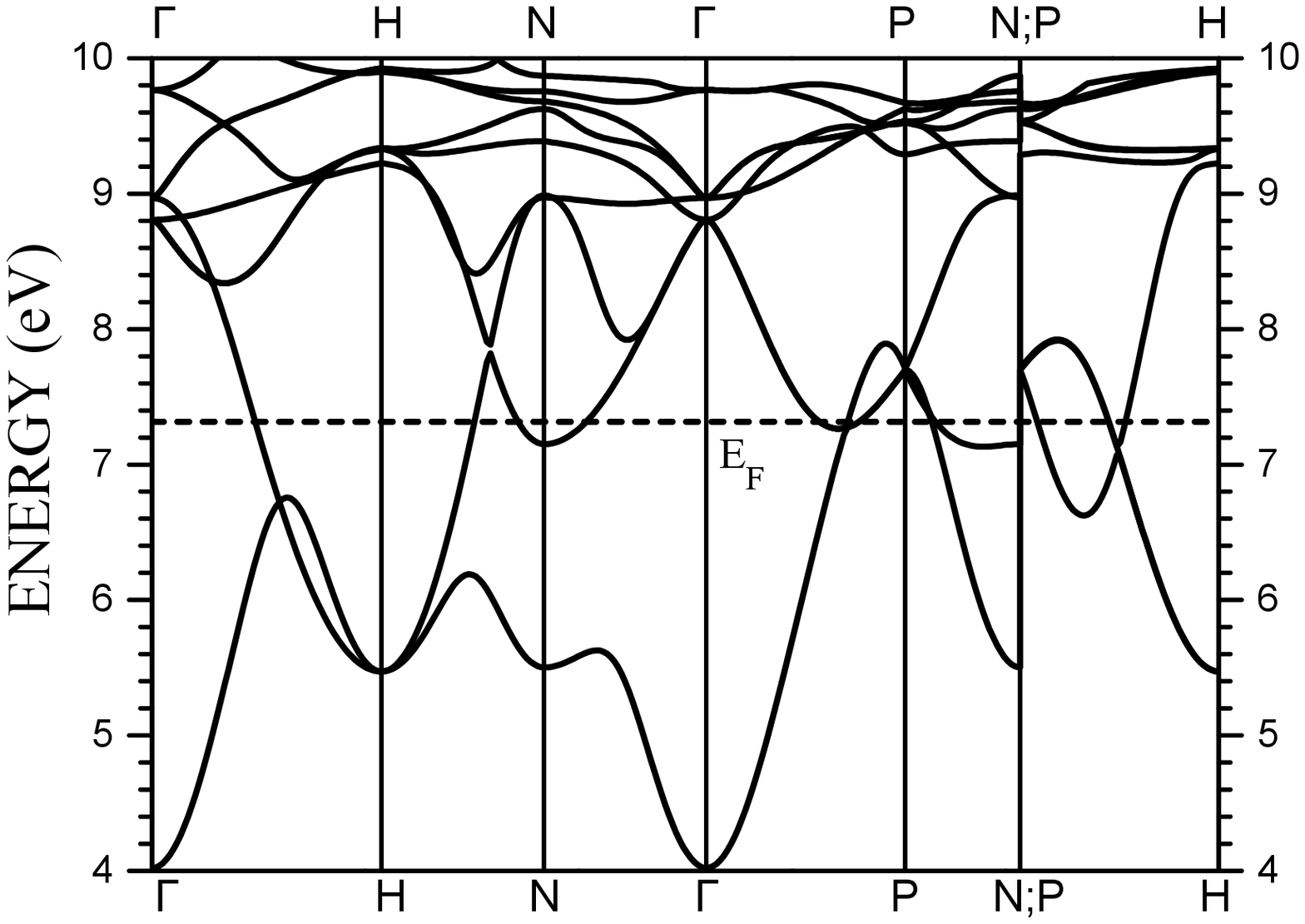}\label{fig2a}}

\subfloat[]{\includegraphics[width=\linewidth]{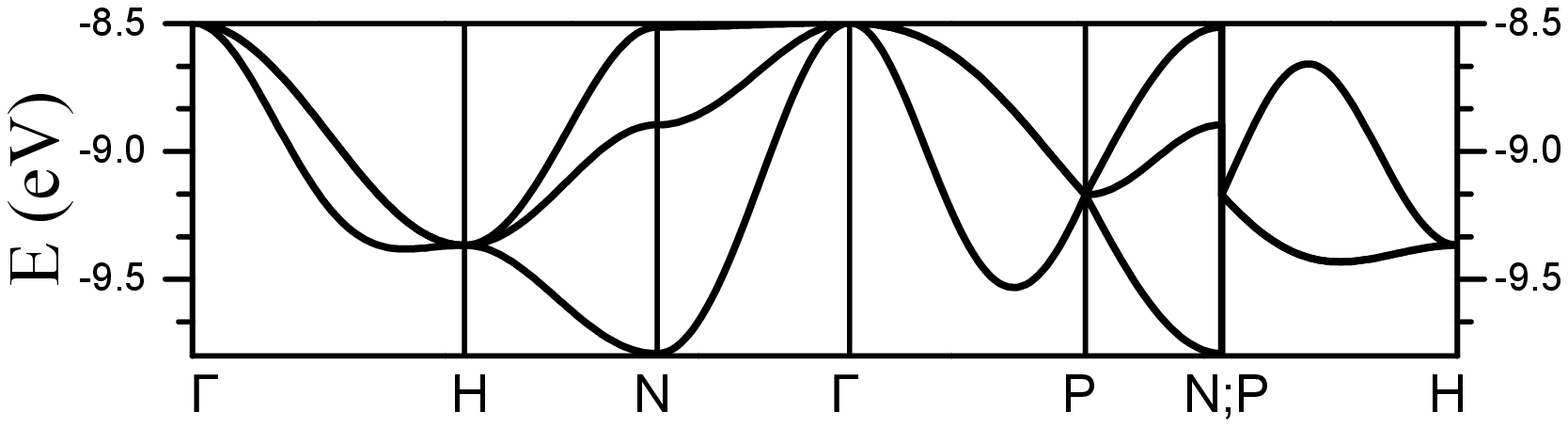}\label{fig2b}}
\caption{ Electronic band structure of bcc $\gamma-$La
along high symmetry lines of the Brillouin zone ($a=4.243$ {\AA}).
(a) valence bands, (b) semicore (5$p$) band.
The horizontal dashed line indicates the Fermi
level.} \label{fig2}
\end{figure}

The equilibrium lattice constant found for bcc-La, $a=4.243$ {\AA},
is in good correspondence with the experimental value $a_{exp}=4.25$~{\AA} \cite{bcc-La}.
The calculated band structure of bcc lattice of lanthanum (PBE exchange and correlation) is plotted in Fig.~\ref{fig2}.

The results of the present and LAPW+LO approaches are compared in
Tables \ref{tab1b}, \ref{tab2b} and~\ref{tab3b}.
%
\begin{table}
\caption{ Total energy ($E_{tot}$, in eV) for various basis sets for bcc calculations of La. $a=4.243$ {\AA},
$R_{MT}=3.355$~{\AA}, $E_0=-231170$ eV, $\triangle E= E_{tot}($FLAPW++$)\,-E_{tot}($FLAPW+).
FLAPW++ stands for the present scheme (FLAPW + 2STBFs) with two radial functions
and FLAPW+ for the FLAPW + LO method with a single radial function.
\label{tab1b} }

\begin{ruledtabular}
\begin{tabular}{c | c  c  c  }

$R_{MT} \cdot K_{max}$ & FLAPW++         & FLAPW+       &  $\triangle E$ \\
\tableline
   7.0                 & $E_0-8.8166$    & $E_0-8.7373$ & -0.0793  \\
   8.0                 & $E_0-9.3567$    & $E_0-9.3428$ & -0.0139  \\
   9.0                 & $E_0-9.4574$    & $E_0-9.4545$ & -0.0029 \\

\end{tabular}
\end{ruledtabular}
\end{table}
%
\begin{table}
\caption{ Energy parameters (in eV) for various basis sets for bcc calculations of La ($a=4.243$ {\AA},
$R_{MT}=3.355$~{\AA}).
FLAPW++ stands for the present scheme (FLAPW + 2STBFs) with two radial functions
and FLAPW+ for the FLAPW + LO method with a single radial function.
\label{tab2b} }

\begin{ruledtabular}
\begin{tabular}{c c | c  c |  c  c }

         &            & \multicolumn{2}{c|}{semicore}    & \multicolumn{2}{c}{valence} \\
         &            & \multicolumn{2}{c|}{$5p-$band}   & \multicolumn{2}{c}{$(spd)-$band} \\
    & $R_{MT} K_{max}$ &   $E_{bot}$  &  $E_{top}$   & $E_{bot}$   &  $E_F$ \\
\tableline
         & 7.0        & -10.7891 & -9.5738 & 3.1271 & 6.5415 \\
 FLAPW++ & 8.0        & -10.2179 & -8.9376 & 3.6086 & 6.9193 \\
         & 9.0        &  -9.7907 & -8.4968 & 4.0213 & 7.3193 \\
\tableline
         & 7.0        & -10.5792 & -9.3654 & 3.2761 & 6.7062 \\
  FLAPW+ & 8.0        & -10.0659 & -8.7832 & 3.7403 & 7.0573 \\
         & 9.0        &  -9.7178 & -8.4227 & 4.0875 & 7.3877 \\

\end{tabular}
\end{ruledtabular}
\end{table}
%
\begin{table}
\caption{ Energy band spectrum (in eV) of bcc La ($a=4.243$ {\AA},
$R_{MT}=3.355$~{\AA}, $R_{MT} \cdot K_{max}=9$) at the $\Gamma$-point
of the Brillouin zone.
FLAPW++ stands for the present scheme (FLAPW + 2STBFs) with two radial functions
and FLAPW+ for the FLAPW + LO method with a single radial function.
\label{tab3b} }

\begin{ruledtabular}
\begin{tabular}{c  c  c  c  }

band & deg. & FLAPW++  & FLAPW+ \\
\tableline
 1    & (3) & -8.4968 & -8.4227  \\
 2    &     &  4.0213 &  4.0875  \\
$E_F$ &     &  7.3193 &  7.3877 \\
 3    & (3) &  8.8077 &  8.8758 \\
 5    & (3) &  8.9685 &  9.0497 \\

\end{tabular}
\end{ruledtabular}
\end{table}
As for fcc-La, the present scheme gives lower values
of total energy of bcc-La for all basis sets. The total energy difference increases with worsening of the basis quality, Table \ref{tab1b}.
For the best basis set ($R_{MT} \cdot K_{max}=9$)
the present method gives energy spectrum shifted downwards by $\approx 0.07$~eV
in comparison with FLAPW+LO band energy values, Tables \ref{tab3b}, \ref{tab2b}.
The difference between energy bands parameters reaches 0.2 eV for poor basis set ($R_{MT} \cdot K_{max}=7$),
Table \ref{tab2b}.
For the best basis set the total energy of bcc-La ($\gamma-$phase) is $\sim0.1$ eV higher than the total energy
of fcc-La ($\beta-$phase), Tables \ref{tab1f} and \ref{tab1b}, which is in agreement with the fact that at normal pressure $\gamma-$La exists only
at high temperatures ($>$1138 K) in narrow temperature range (53 K) \cite{bcc-La}.

\subsection{Hexagonal close packed structure of Cd}
\label{app-Cd}

\begin{figure}[t]
\includegraphics[width=\linewidth]{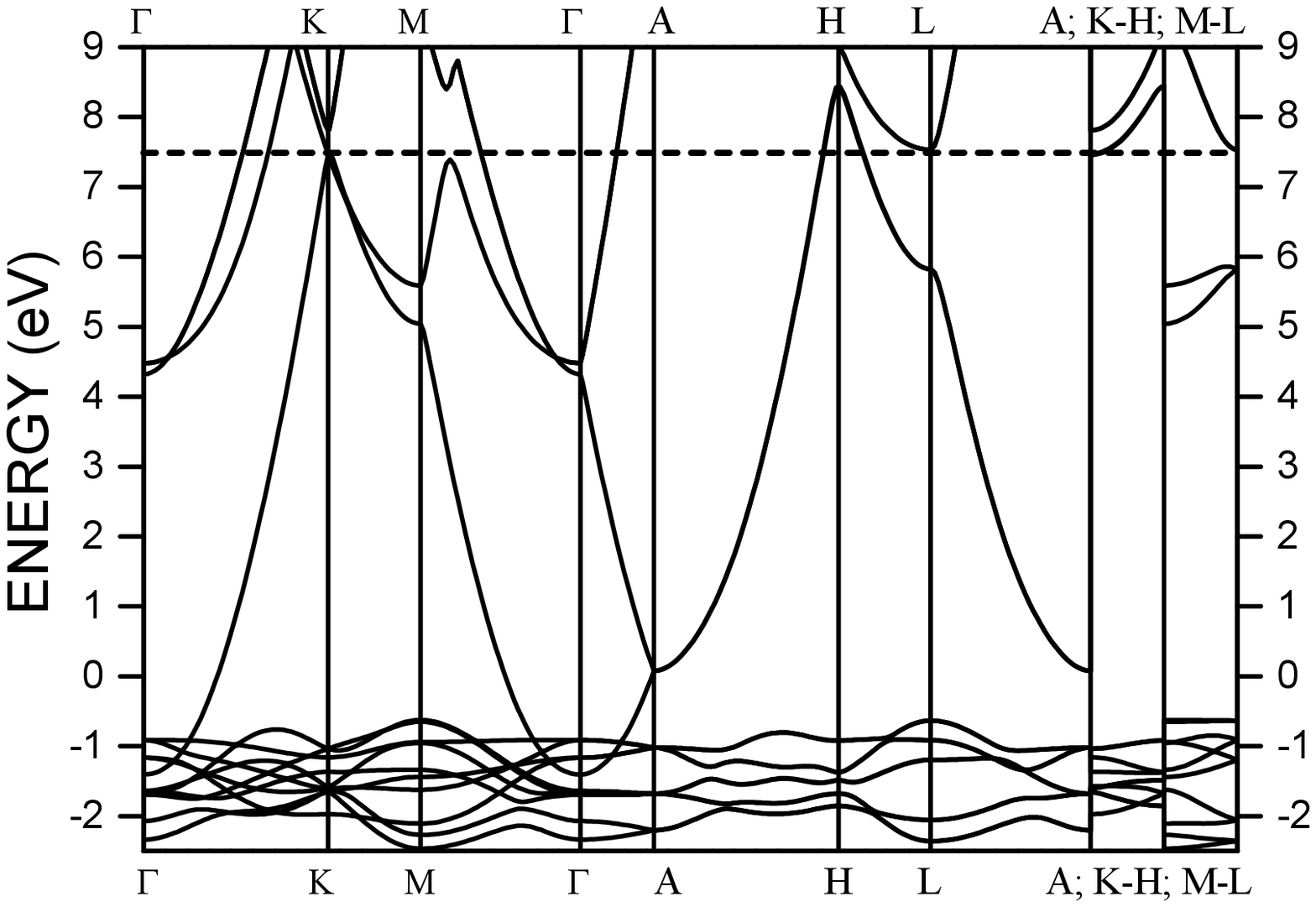}

\caption{ (a) Electronic band structure of hexagonal close packed
Cd along high symmetry lines of the Brillouin zone ($a=2.986$~{\AA}, $c=5.632$~{\AA}). }
\label{fig4}
\end{figure}

Hexagonal close packed (hcp) cadmium is a special case because its
$4d-$states are not separated from the valence $5s-$states by an
energy gap, Fig.\ \ref{fig4}.
In fact there is a small overlap between the top of $d-$band and
the bottom of $s-$band. This implies that if the linear expansion
energy of extended $d-$states is properly chosen (for example,
at 0.5 eV above the $d-$band bottom energy, $E(d)=-1.97$~eV) the electronic band structure of cadmium can be
carried out without supplemented tight-binding basis states of
$d-$type. Thus, both calculations i.e. with and without
supplemented states, can be directly compared with each other. The
second peculiarity is that because of the gapless energy spectrum
the radial distribution of $d-$states does not change much
throughout the valence band. If we chose the linear expansion
energy of supplemented $d-$states
at 1.0 eV below the Fermi
energy ($E_s(d)=6.42$~eV), then the energy difference $E_s(d)-E_e(d)
\sim 8.5$~eV is a relatively small value. In that case one
can expect that the extended $d-$states and the supplemented
$d-$states are close to being linearly dependent, and we can test
the scenario described in Appendix~\ref{appD}.
(Two supplementary radial functions, Eqs.\ (\ref{p1}) and (\ref{p2}), are specified by the
coefficients $a_{s,1}=-0.3448$, $b_{s,1}=0.4643$ for $R_1(r)$ and
$a_{s,2}=-1.9343$, $b_{s,2}=-0.2952$ for $R_2(r)$.)

The technical parameters of FLAPW calculations were the following:
the plane wave cut off parameter $K_{max} R_{MT} < 9$ (149 basis
states and 10 supplemented tight binding $d-$states), $L_{max}=8$
for the expansion of electron density and wave functions inside
the MT-spheres, 216 special points in the irreducible
part of the Brillouin zone during the self-consistent procedure, and the PBE \cite {PBE}
form of the exchange correlation potential.

The equilibrium lattice constants are $a=2.986$~{\AA}, $c=5.632$~{\AA},
which are in good correspondence with the experimental values,
$a_{exp}=2.9794$~{\AA} and $c_{exp}=5.6186$~{\AA}  \cite{Edw-Cd}.
Band energies are shown in Fig.~\ref{fig4}.

Comparison between the present approach, the LAPW+LO and the standard LAPW treatment is presented
in Tables \ref{tab1c}, \ref{tab2c} and \ref{tab3c}.
%
\begin{table}
\caption{ Total energy ($E_{tot}$, in eV) for various basis sets for hcp calculations of Cd. $a=2.986$~{\AA}, $c=5.632$~{\AA},
$R_{MT}=2.74$~{\AA}, $E_0=-304510$ eV, $\triangle E= E_{tot}($FLAPW++$)\,-E_{tot}($FLAPW+).
FLAPW++ stands for the present scheme (FLAPW + 2STBFs) with two radial functions
and FLAPW+ for the FLAPW + LO method with a single radial function.
\label{tab1c} }

\begin{ruledtabular}
\begin{tabular}{c | c  c  c  c }
$R_{MT} K_{max}$ & FLAPW++         & FLAPW+       & $\triangle E$ & FLAPW   \\
\tableline
   7.0                 & $E_0-4.3780$    & $E_0-3.8336$ & -0.5444 & $E_0-2.4724$  \\
   7.5                 & $E_0-6.6012$    & $E_0-6.4998$ & -0.1014 & $E_0-6.3540$  \\
   8.0                 & $E_0-7.4471$    & $E_0-7.4128$ & -0.0343 & $E_0-7.3592$  \\
   8.5                 & $E_0-7.9324$    & $E_0-7.9033$ & -0.0291 & $E_0-7.8428$  \\
   9.0                 & $E_0-8.0093$    & $E_0-7.9899$ & -0.0194 & $E_0-7.9459$  \\

\end{tabular}
\end{ruledtabular}
\end{table}
%
\begin{table}
\caption{ Energy parameters (in eV) for various basis sets for hcp calculations of Cd.
FLAPW++ stands for the present scheme (FLAPW + 2STBFs) with two radial functions
and FLAPW+ for the FLAPW + LO method with a single radial function.
\label{tab2c} }

\begin{ruledtabular}
\begin{tabular}{c c |  c  c }

  & $R_{MT} \cdot K_{max}$  &   $E_{bot}$  &  $E_F$ \\
\tableline
          & 7.0   & -3.1660 & 6.8094 \\
          & 7.5   & -2.9380 & 7.1192 \\
  FLAPW++ & 8.0   & -2.7794 & 7.2053 \\
          & 8.5   & -2.5564 & 7.3358 \\
          & 9.0   & -2.4655 & 7.4186 \\
\tableline
           & 7.0  & -2.9447 & 6.9453 \\
           & 7.5  & -2.8216 & 7.2016 \\
  FLAPW+   & 8.0  & -2.6999 & 7.2648 \\
           & 8.5  & -2.4763 & 7.3969 \\
           & 9.0  & -2.3949 & 7.4741 \\
\tableline
           & 7.0  & -3.3958 & 7.0200 \\
           & 7.5  & -2.9187 & 7.2652 \\
   FLAPW   & 8.0  & -2.7556 & 7.2997 \\
           & 8.5  & -2.5211 & 7.4394 \\
           & 9.0  & -2.4273 & 7.5123 \\

\end{tabular}
\end{ruledtabular}
\end{table}
%
\begin{table}
\caption{
Energy band spectrum (in eV) of hcp structure of Cd at the $\Gamma-$point of the Brillouin zone ($a=2.986$~{\AA}, $c=5.632$~{\AA})
with PBE exchange \cite{PBE} ($E_F$ is the Fermi energy, deg.\ is the energy degeneracy).
FLAPW++ stands for the present scheme (FLAPW + 2STBFs) with two radial functions
and FLAPW+ for the FLAPW + LO method with a single radial function.
\label{tab3c} }

\begin{ruledtabular}
\begin{tabular}{l c c c c }
      & deg. & FLAPW++ & FLAPW+ & FLAPW \\
\tableline
  1 &     & -2.3356 & -2.2662 & -2.2954 \\
  2 &     & -2.0708 & -2.0009 & -2.0305 \\
  3 & (2) & -1.6919 & -1.6194 & -1.6604 \\
  4 & (2) & -1.6399 & -1.5668 & -1.6078 \\
  5 &     & -1.4047 & -1.3485 & -1.3215 \\
  6 & (2) & -1.1638 & -1.0888 & -1.1334 \\
  7 & (2) & -0.9125 & -0.8377 & -0.8866 \\
  8 &     &  4.3215 &  4.3750 &  4.4176 \\
  9 &     &  4.4792 &  4.5374 &  4.5479 \\
$E_F$ &   &  7.4186 &  7.4741 &  7.5123 \\
 10 &     & 17.6841 & 17.7375 & 17.7711 \\

\end{tabular}
\end{ruledtabular}
\end{table}
We observe that for all basis sets the present approach gives lower total energy values, Table \ref{tab1c}.
For the best basis set ($R_{MT} K_{max} = 9$) the obtained total energy difference with the LAPW+LO value, 0.02 eV, is larger than for fcc-La or bcc-La.
In the present approach band energies computed with the best basis set are lowered by 0.04-0.09 eV in comparison with LAPW+LO and LAPW values,
Tables \ref{tab2c} and \ref{tab3c}.
These energy shifts are also typical for the other points of the Brillouin zone.

\section{CONCLUSIONS}
\label{Con}

We have presented a new method for the improvement of the LAPW
description of the electronic band structure by using two linearization energies
for the same $(l,m)$ partial component.
Starting with two LAPW radial
functions, Eqs.\ (\ref{i4a}) and (\ref{i4b}), having the same
angular dependence $Y_{l_c,m}(\hat{r})$ but different linearization energies ($E_l^{(1)}$ and $E_l^{(2)}$)
inside MT-spheres, we have demonstrated that their augmentation to the basis plane wave can be performed by
constructing additional basis functions $\phi_{s,i}$ ($i=1,2$) in
the form of Eq.\ (\ref{m11}) [two functions, Eqs.\ (\ref{m10b})
and (\ref{m10c}), for each $l,m$-component].
The supplementary basis functions have zero values
and slopes on the sphere surface, Eqs.\ (\ref{m12a}),
(\ref{m12b}), and are linear independent of the
usual LAPW basis states. The constructed basis functions are of the
tight-binding type, Eq.\ (\ref{m16}), and obey Bloch's law.

In contrast to the LAPW+LO method with only one supplemented function, Eq.\ (\ref{m10b}), in our treatment for each
$l,m$-component there are two supplemented functions
[Eq.\ (\ref{m10b}) and (\ref{m10c})].
The second basis function (absent in LAPW+LO) closely examined in this work,
owes its appearance to the $\dot{u}_l$ function in the canonical LAPW method.
Thus, the basis sets of LAPW and LAPW+LO methods can be extended further by adding supplemented functions
of the tight-binding type, Eq.\ (\ref{m10c}).

In Sec.\ \ref{App}, the present method with extended basis set has been applied to the study of the face centered
and body centered phases of lanthanum ($\beta-$La and $\gamma-$La)
with the $5p-$semicore shell separated by a gap of forbidden states from the valence states and
to the hexagonal close packed structure of cadmium, where the
semicore $4d-$states overlap with the valence $5s-$states.
In all cases we have observed a systematic improvement
in the values of total energy in comparison with the standard LAPW+LO
treatment, Tables \ref{tab1f}, \ref{tab1b}, \ref{tab1c}.
The difference with LAPW+LO total energy is only 0.003-0.004 eV for La and 0.019 eV for Cd for the best basis set
($R_{MT} \cdot K_{max}=9$) but significantly increases in going to intermediate
($R_{MT} \cdot K_{max}=8.5$ or 8) and poor ($R_{MT} \cdot K_{max}=7.5$ or 7) basis sets.

\acknowledgments

A.V.N. acknowledges useful discussions with
B. Verberck, E.V. Tkalya, A.V. Bibikov.

\appendix

\section{}
\label{appA}

Solution to the system of linear equations, Eq.\ (\ref{m7a}), is
given by
\begin{subequations}
\begin{eqnarray}
  a_{s,1} = \frac{1}{\triangle} ( u_e \dot{u}'_s - u'_e \dot{u}_s ) , \label{a1a} \\
  b_{s,1} = \frac{1}{\triangle} ( u'_e u_s - u_e u'_s ) , \label{a1b}
\end{eqnarray}
\end{subequations}
and the solution to the system (\ref{m7b}) is
\begin{subequations}
\begin{eqnarray}
  a_{s,2} = \frac{1}{\triangle} ( \dot{u}_e \dot{u}'_s - \dot{u}'_e \dot{u}_s ) , \label{a2a} \\
  b_{s,2} = \frac{1}{\triangle} ( \dot{u}'_e u_s - \dot{u}_e u'_s ) . \label{a2b}
\end{eqnarray}
\end{subequations}
Here
\begin{eqnarray}
  \triangle =   \dot{u}_s u'_s - u_s \dot{u}'_s  \approx \frac{1}{R_{MT}^2}  . \label{a3}
\end{eqnarray}

\section{}
\label{appB}

The matrix elements for the overlap and Hamiltonian operator
between supplementary states,
\begin{subequations}
\begin{eqnarray}
 \langle \phi_{s,i} | O | \phi_{s,j} \rangle  = O_s^i{}_s^j , \label{b0a} \\
 \langle \phi_{s,i} | H | \phi_{s,j} \rangle  = H_s^i{}_s^j , \label{b0b}
\end{eqnarray}
\end{subequations}
are partitioned in three different blocks, when $i=j=1$ (block $I$), $i=j=2$ (block $II$),
and $i=1$, $j=2$ or $i=2$, $j=1$ (block $III$).

For the first block ($I$) we have
\begin{subequations}
\begin{eqnarray}
  O_s^1{}_s^1(\vec{k}) = {\cal O}_{s,s} \;
  (1 + C_{s}^1 {}_{s}^1 + C_e {}_s^1 ) ,
  \label{b1a}
\end{eqnarray}
where $N_{\alpha}$ is the number of equivalent spheres $\alpha$, while
\begin{eqnarray}
  & & {\cal O}_{s,s} = \frac{(4\pi)^2}{V}(R^{\alpha}_{MT})^4\, N_{\alpha} , \label{b1b} \\
  & & C_{s}^1 {}_{s}^1 = a_{s,1} a_{s,1} + b_{s,1} b_{s,1} \, {\cal N}(\dot{u}_s, \dot{u}_s)  , \label{b1c} \\
  & & C_e {}_s^1 = a_{s,1}{\cal N}(u_e, u_s) + b_{s,1}\, {\cal N}(u_e, \dot{u}_s)  , \label{b1d}
\end{eqnarray}
\end{subequations}
and ${\cal N}$ stands for the integral over the product of two
functions,
\begin{eqnarray}
   {\cal N}(u_1, u_2) = \int_{0}^{R_{MT}^{\alpha}} u_1(r)\, u_2(r)\, r^2 dr  . \label{b2}
\end{eqnarray}
Notice that ${\cal N}(u_e, u_e)={\cal N}(u_s, u_s)=1$ and ${\cal
N}(u_e, \dot{u}_e)={\cal N}(u_s, \dot{u}_s)=0$. It is also assumed
here that the constant coefficient of the supplementary function
is taken in the form of Eq.\ (\ref{i17}). If another form is used,
the factor ${\cal O}_{s,s}$, Eq.\ (\ref{b1b}), [and ${\cal
O}_{e,s}(\vec{k}_n)$, Eq.\ (\ref{c2c})] should be changed
accordingly.

For the block $II$ we obtain
\begin{subequations}
\begin{eqnarray}
  O_s^2{}_s^2 = {\cal O}_{s,s} \,
  ( {\cal N}(\dot{u}_e, \dot{u}_e) + C_{s}^2 {}_{s}^2 + C_e {}_s^2 ) ,
  \label{b3a}
\end{eqnarray}
where
\begin{eqnarray}
  C_{s}^2 {}_{s}^2 = a_{s,2} a_{s,2} + b_{s,2} b_{s,2} \, {\cal N}(\dot{u}_s, \dot{u}_s)  , \label{b3b} \\
  C_e {}_s^2 = a_{s,2}{\cal N}(\dot{u}_e, u_s) + b_{s,2}\, {\cal N}(\dot{u}_e, \dot{u}_s)  . \label{b3c}
\end{eqnarray}
\end{subequations}

Finally, for the block $III$ we get
\begin{subequations}
\begin{eqnarray}
  O_s^1{}_s^2 = {\cal O}_{s,s} \,
  (  C_{s}^1 {}_{s}^2 + C_e {}_s^1 + C_e {}_s^2 ) ,
  \label{b4a}
\end{eqnarray}
where
\begin{eqnarray}
  C_{s}^1 {}_{s}^2 = a_{s,1} a_{s,2} + b_{s,1} b_{s,2} \, {\cal N}(\dot{u}_s, \dot{u}_s)  , \label{b5b}
\end{eqnarray}
\end{subequations}
and $C_e {}_s^1$ is given by Eq.\ (\ref{b1d}), while $C_e {}_s^2$ by Eq.\ (\ref{b3c}).

For the matrix elements of the Hamiltonian $H_s^i{}_s^j(\vec{k})$,
Eq.\ (\ref{b0b}), we also obtain three blocks. For the first block
($I$) we have
\begin{eqnarray}
  H_s^1{}_s^1 = {\cal O}_{s,s} \,
  \left( E_e + E_s\, C_{s}^1 {}_{s}^1 + a_{s,1} b_{s,1} \right.   \nonumber \\
  + \left. (E_e + E_s)\, C_e {}_s^1 + b_{s,1} {\cal N}(u_e,u_s) \right) .
  \label{b6}
\end{eqnarray}
Here $E_e$ and $E_s$ are energies at which the radial wave functions $u_e(r)$ and $u_s(r)$
are evaluated in the $MT$-sphere $\alpha$.

For the second block ($II$) we get
\begin{subequations}
\begin{eqnarray}
  H_s^2{}_s^2 = {\cal O}_{s,s} \,
  \left( E_e\, {\cal N}(\dot{u}_e,\dot{u}_e) + E_s\, C_{s}^2 {}_{s}^2 + a_{s,2} b_{s,2} \right.   \nonumber \\
  + \left. (E_e + E_s)\, C_e {}_s^2 + \gamma_e {}_s^2 \right) , \quad   \quad
  \label{b7a}
\end{eqnarray}
where
\begin{eqnarray}
\gamma_e {}_s^2 = a_{s,2}\, {\cal N}(u_e, u_s) + b_{s,2}\, \left( {\cal N}(u_e,\dot{u}_s) + {\cal N}(\dot{u}_e,u_s) \right) . \nonumber \\
  \label{b7b}
\end{eqnarray}
\end{subequations}

For the third block ($III$) we have
\begin{eqnarray}
  H_s^1{}_s^2 = {\cal O}_{s,s} \,
  \left( E_s\, ( C_{s}^1 {}_{s}^2 + C_e {}_s^1 ) + a_{s,2} b_{s,1} \right.   \nonumber \\
  + \left.  E_e\, C_e {}_s^2 + b_{s,1}\, {\cal N}(\dot{u}_e, u_s) \right) . \quad   \quad
  \label{b8}
\end{eqnarray}

\section{}
\label{appC}

In this section we quote the expressions for matrix elements for
the overlap and Hamiltonian operator between supplementary and
extended states,
\begin{subequations}
\begin{eqnarray}
 \langle \phi_{s,i} | O | \phi_{n} \rangle  = O_{s,}^{i,}{}_e^n(\vec{k}) , \label{c1a} \\
 \langle \phi_{s,i} | H | \phi_{n} \rangle  = H_{s,}^{i,}{}_e^n(\vec{k}) . \label{c1b}
\end{eqnarray}
\end{subequations}
The extended states here are the usual LAPW basis functions, Eq.\ (\ref{i1}),
which are characterized by the wave vector $\vec{k}_n=\vec{k}+\vec{K}_n$.
The supplementary functions have two components $i=1,2$, for
each $l,m$-angular dependence, Eq.\ (\ref{m10b}), (\ref{m10c}), (\ref{m11}).

For the matrix of overlap we get
\begin{subequations}
\begin{eqnarray}
  O_s^1{}_e^n(\vec{k}) = {\cal O}_{e,s}(\vec{k}_n) \, S_s^1{}_e^n , \label{c2a}
\end{eqnarray}
where
\begin{eqnarray}
 S_s^1{}_e^n =
   a_e + a_{s,1} a_e\, {\cal N}(u_e, u_s) + a_{s,1} b_e\, {\cal N}(\dot{u}_e, u_s)  \nonumber \\
   + b_{s,1} a_e\, {\cal N}(u_e, \dot{u}_s) + b_e b_{s,1}\, {\cal N}(\dot{u}_e, \dot{u}_s) ,   \label{c2b}
\end{eqnarray}
and the structure factor is
\begin{eqnarray}
  {\cal O}_{e,s}(\vec{k}_n) = \frac{(4\pi)^2}{V}(R^{\alpha}_{MT})^4\, Y_{lm}^*(\hat{k}_n)  , \nonumber \\
  \times \sum_{\nu} exp[i(\vec{k}_n - \vec{k})\, \vec{r}_{\nu,\alpha}] . \label{c2c}
\end{eqnarray}
\end{subequations}
Here, $\vec{r}_{\nu,\alpha}$ stands for the coordinates of all $\nu$ centers of $MT$-spheres of the type $\alpha$ in the primitive unit cell.
The factors ${\cal N}(u_1, u_2)$ in (\ref{c2b}) are integrals between two functions given by Eq.\ (\ref{b2}).

The quantities $a_e$ and $b_e$ in Eq.\ (\ref{c2b}) and below are
standard LAPW expansion coefficients for the component with $l,m$,
defined by Eq.\ (\ref{m4a}) and (\ref{m4b}), i.e. $a_e=a_e^0$ and
$b_e=b_e^0$. Since they depend on $\vec{k}_n$, $l$, and $\alpha$, we can write
$a_e=a_l^{\alpha}(\vec{k})$, $b_e=b_l^{\alpha}(\vec{k})$. Explicit
expressions for them can be found in Ref.\ \cite{Koe,blapw}.

The matrix element for the second case ($i=2$) reads as
\begin{subequations}
\begin{eqnarray}
  O_s^2{}_e^n(\vec{k}) = {\cal O}_{e,s}(\vec{k}_n) \, S_s^2{}_e^n ,  \label{c3a}
\end{eqnarray}
where
\begin{eqnarray}
  S_s^2{}_e^n &=&
   b_e\, {\cal N}( \dot{u}_e, \dot{u}_e)  + a_{s,2} a_e\, {\cal N}(u_e, u_s) + a_{s,2} b_e\, {\cal N}(\dot{u}_e, u_s)  \nonumber \\
   & & + b_{s,2} a_e\, {\cal N}(u_e, \dot{u}_s) + b_e b_{s,2}\, {\cal N}(\dot{u}_e, \dot{u}_s) .    \label{c3b}
\end{eqnarray}
\end{subequations}

Below we quote the matrix elements for the Hamiltonian,
\begin{eqnarray}
  H_s^1{}_e^n(\vec{k}) = {\cal O}_{e,s}(\vec{k}_n)\,  (  E_e\, S_s^1{}_e^n + b_e + a_{s,1} b_e\, {\cal N}(u_e, u_s)  \nonumber \\
                       + b_{s,1} b_e\, {\cal N}(u_e, \dot{u}_s) ) , \quad \quad
  \label{c4}   \\
  H_s^2{}_e^n(\vec{k}) = {\cal O}_{e,s}(\vec{k}_n)\,  (  E_e\, S_s^2{}_e^n + a_{s,2} b_e\, {\cal N}(u_e, u_s)    \nonumber \\
                       + b_{s,2} b_e\, {\cal N}(u_e, \dot{u}_s) ) .  \quad \quad
  \label{c5}
\end{eqnarray}
Here ${\cal O}_{e,s}$ is given by Eq.\ (\ref{c2c}), while $S_s^i{}_e^n$ by Eq.\ (\ref{c2b}) for $i=1$, and by Eq.\ (\ref{c3b}) for $i=2$.

\section{}
\label{appD}

Here we demonstrate that the two supplementary basis functions
$\phi_{s,i}$ ($i=1,2$), Eq.\ (\ref{m11}), work even in the case
when the expansion energies $E_s$ and $E_e$ lie not far from each other.

Consider $E_e = E_s + \varepsilon $, where $\varepsilon/E_s \ll 1$. Making use of the following expansions
\begin{subequations}
\begin{eqnarray}
   u_e(r) &=& u_s(r) + \dot{u}_s(r) \, \varepsilon   \nonumber \\
   & &+ \frac{1}{2} \ddot{u}_s(r) \, \varepsilon^2 + \frac{1}{6} \dddot{u}_s(r) \, \varepsilon^3 + O(\varepsilon^4) ,  \label{m17a} \\
   \dot{u}_e(r) &=& \dot{u}_s(r) + \ddot{u}_s(r) \, \varepsilon + \frac{1}{2} \dddot{u}_s(r) \, \varepsilon^2 + O(\varepsilon^3) , \nonumber \\  \label{m17b}
\end{eqnarray}
\end{subequations}
 for $u_e(r)$ and $\dot{u}_e(r)$ and substituting them in Eqs.\ (\ref{m10b}), (\ref{m10c}), we arrive at
 \begin{subequations}
\begin{eqnarray}
   R_{l,m}^{s,1}(r) = \frac{1}{2} \varepsilon^2 \, C_0\, e^{i \vec{k} \vec{R}_{\alpha}}\,
   [ \ddot{u}_s + \frac{1}{3} \varepsilon\, \dddot{u}_s + O(\varepsilon^2)  \nonumber \\
   + a'_{s,1}\, u_s + b'_{s,1}\, \dot{u}_s ], \label{m18a} \\
   R_{l,m}^{s,2}(r) = \varepsilon \, C_0\, e^{i \vec{k} \vec{R}_{\alpha}}\,
   [ \ddot{u}_s + \frac{1}{2} \varepsilon \, \dddot{u}_s + O(\varepsilon^2)  \nonumber \\
   + a'_{s,2}\, u_s + b'_{s,2}\, \dot{u}_s ] . \label{m18b}
\end{eqnarray}
\end{subequations}
Here
\begin{subequations}
\begin{eqnarray}
   a'_{s,1} = \frac{1}{\varepsilon^2}(a_{s,1} + 1),  \quad \quad b'_{s,1} = \frac{1}{\varepsilon^2}(b_{s,1} + \varepsilon );
   \label{m19a} \\
   a'_{s,2} = \frac{1}{\varepsilon}\, a_{s,2},  \quad \quad \quad b'_{s,2} = \frac{1}{\varepsilon}\,(b_{s,2} + 1 ) .  \label{m19b}
\end{eqnarray}
\end{subequations}
The prefactors $\varepsilon^2/2$ and $\varepsilon$ in Eqs.\
(\ref{m18a}), (\ref{m18b}) are not very important, because as a
consequence of solving secular equations these basis functions
will be effectively orthonormalized. Functions $R_{l,m}^{s,1}(r)$
and $R_{l,m}^{s,2}(r)$ have two important features. First, by a
linear transformation the two functions, Eqs.\ (\ref{m18a}),
(\ref{m18b}), can be transformed to two functions with the
following radial dependencies
\begin{subequations}
\begin{eqnarray}
   U_1(r) = \ddot{u}_s + a''_{s,1}\, u_s + b''_{s,1}\, \dot{u}_s ,   \label{m20a} \\
   U_2(r) = \dddot{u}_s + a''_{s,1}\, u_s + b''_{s,1}\, \dot{u}_s ,  \label{m20b}
\end{eqnarray}
\end{subequations}
where the coefficients $a''_{s,1}$ etc. can be expressed through
$a'_{s,1}$ etc. Since the functions $\ddot{u}_s(r)$ and
$\dddot{u}_s(r)$ are linear independent, the same property applies
to $U_1(r)$ and $U_2(r)$ and, consequently, to $R_{l,m}^{s,1}(r)$
and $R_{l,m}^{s,2}(r)$, although the property deteriorates as
$\varepsilon \rightarrow 0$. Second, we still can impose the
boundary conditions (\ref{m12a}) and (\ref{m12b}).

However, since the initial functions $R_{l,m}^{s,1}(r)$ and
$R_{l,m}^{s,2}(r)$, have almost identical radial dependence
(neglecting terms with $\varepsilon\, \dddot{u}_s$ and the others
of high order of $\varepsilon$ in Eqs.\ (\ref{m18a}), (\ref{m18b})
make them completely identical), the normalization procedure leads
to the appearance of an effective basis state $\phi'$, which is
orthogonal to other basis states and expressed through the linear
combination of initial states, $\phi'=C_1\, \psi_{s,1} + C_2\,
\psi_{s,2}$ with large coefficients $C_1$ and $C_2$ (i.e.
$|C_1| \gg 1$, $|C_2| \gg 1$). In that case, the partial charges
of supplementary basis states can also be very large.
Nevertheless, some of these partial charges are of opposite sign
and effectively cancel each other in the final answer. We have
observed the effect in the calculation of the hexagonal close
packed lattice of cadmium reported in Sec.\ \ref{app-Cd}.



\begin{references}

\bibitem{Sza} A. Szabo and N. S. Ostlund, {\it Modern Quantum Chemistry} (Dover, New York, 1996).
\bibitem{Dun} T. H. Dunning, Jr., J. Chem. Phys. {\bf 90}, 1007 (1989).
\bibitem{And} O.K. Andersen, Phys. Rev. B {\bf 12}, 3060 (1975).
\bibitem{Koe} D.D. Koelling and G.O. Arbman, J. Phys. F {\bf 5}, 2041 (1975).
\bibitem{blapw} D.J. Singh, L. Nordstr\"{o}m, {\it Planewaves, Pseudopotentials, and the LAPW Method}, 2nd ed. (Springer, New York, 2006).
\bibitem{wien2k} P. Blaha, K. Schwarz, G. Madsen, D. Kvasnicka and J. Luitz,
J. Luitz, WIEN2K: {\it An Augmented Plane Wave plus Local
Orbitals Program for Calculating Crystal Properties}
(Vienna University of Technology, Austria, 2001).
\bibitem{Sin-2} D. Singh, Phys. Rev. B {\bf 43}, 6388 (1991).
\bibitem{Mat} L.F Mattheiss and D.R. Hamann, Phys. Rev B {\bf 33}, 823 (1986).
\bibitem{Sin-1} D. Singh and H. Krakauer, Phys. Rev. B {\bf 43}, 1441 (1991).
\bibitem{Sjo} E. Sj\"{o}stedt, L. Nordstr\"{o}m and D.J. Singh, Solid State Commun. {\bf 114}, 15 (2000).

\bibitem{Bla-1} P. Blaha, D.J. Singh, P.I. Sorantin and K. Schwarz, Phys. Rev. B {\bf 46}, 1321 (1992).

\bibitem{Sin-3} D.J. Singh, K. Schwarz and P Blaha, Phys. Rev. B {\bf 46}, 5849 (1992).

\bibitem{Sin-4} D.J. Singh, Phys. Rev. B {\bf 44}, 7451 (1991).

\bibitem{Goe} S. Goedecker and K. Maschke, Phys. Rev. B {\bf 42}, 8858 (1990).


\bibitem{PBE} J. P. Perdew, K. Burke, and M. Ernzerhof, Phys. Rev. Lett. {\bf 77}, 3865 (1996).

\bibitem{Nik} The FLAPW-Moscow code [registration number 2015616990 (Russia) from 26/06/2015], see also
A.V. Nikolaev, I.T. Zuraeva, G.V. Ionova, and B.V. Andreev, Phys. Solid State {\bf 35}, 213 (1993).

\bibitem{fcc-La}
K. Syassen and W.B. Holzapfel, Solid State Commun. {\bf 16}, 553 (1975).

\bibitem{bcc-La}
F. G\"uthoff, W. Petry, C. Stassis, A. Heiming, B. Hennion, C. Herzig, J. Trampenau,
Phys. Rev. B {\bf 47}, 2563 (1993).

\bibitem{Edw-Cd}
D. A. Edwards, W. E. Wallace, and R. S. Craig, J. Am. Chem. Soc. {\bf 74}, 5256 1952.

\end{references}
\end{document}